# The organization and management of the Virtual Astronomical Observatory

G. Bruce Berriman[a,], Robert J. Hanisch[b,c], T. Joseph W. Lazio[d],
Alexander Szalay[e], Giussepina Fabbiano[f]

[a]Infrared Processing and Analysis Center, California Institute of Technology, 770 South Wilson Ave., Pasadena, CA USA 91125 USA; [b]U.S. Virtual Astronomical Observatory, 1400 16th Street NW, Suite 730, Washington, DC 20036 USA; [c]Space Telescope Science Institute, 3700 San Martin Drive, Baltimore, MD USA 21218; [d]Jet Propulsion Laboratory, California Institute of Technology, M/S 138-308, 4800 Oak Grove Dr., Pasadena, CA 91106 USA; [e]The Johns Hopkins University, Department of Physics and Astronomy, 3400 North Charles St., Baltimore, MD 21211 USA; [f]Smithsonian Astrophysical Observatory, 60 Garden St., Cambridge, MA 02138

## ABSTRACT

The U.S. Virtual Astronomical Observatory (VAO; http://www.us-vao.org/) has been in operation since May 2010. Its goal is to enable new science through efficient integration of distributed multi-wavelength data. This paper describes the management and organization of the VAO, and emphasizes the techniques used to ensure efficiency in a distributed organization. Management methods include using an annual program plan as the basis for establishing contracts with member organizations, regular communication, and monitoring of processes.

## 1. INTRODUCTION

Astronomy is being transformed by the vast quantities of observational data, models, and simulations that are becoming available to astronomers at an ever-accelerating rate, and astronomy has been an international pioneer of data-enabled science. The continuing increase in the rate of data acquisition is expected to continue as new facilities—across the electromagnetic wavelength spectrum and expanding into the time domain and multi-messenger observations—enter operation over the next decade and beyond. Not only is there a growing need for new data standards and methods for access, but also for advanced statistical analysis such as knowledge discovery and data mining.

The Virtual Observatory (VO) is an international concept for enabling access to data within an integrated and seamless astronomical data "ecosystem," by a combination of developing international standards and providing for access. The US Virtual Astronomical Observatory (VAO) is the operational successor to the National Virtual Observatory (NVO) technology and infrastructure development project. The NVO was funded from 2002 to 2008 by the National Science Foundation's Information Technology Research program. The VAO is co-sponsored by the National Science Foundation (Division of Astronomical Sciences) and the National Aeronautics and Space Administration (Astrophysics Division). Funding for the VAO began in May 2010 and is planned to continue for five years. The NVO was a founding member of the International Virtual Observatory Alliance (IVOA), and the VAO continues to play an active role in international standards development activities.

The VAO, building on prior national and international VO infrastructure developments, is providing new and more efficient ways to share, correlate, and integrate data from large surveys, heterogeneous sets of pointed observations, theoretical models and simulations, and scholarly publications. The VAO provides the components, libraries, and templates that allow National facilities, major projects, and end-users to craft their own VO-enabled applications for seamless data access and integration, especially in support of data intensive research. The VAO develops and supports robust, integrated, general-purpose applications that can be accessed through a Web browser or can be downloaded for local installation. Through a combination of VAO-provided tools, community-created applications, software provided by the VAO's international partners, and industry collaborations, the VO is becoming an essential component of the information backbone for the U.S. astronomical research community. The VAO is also augmenting archive-based scientific research (an area pioneered by the NASA data centers and NSF telescopes, but now widely adopted), which is being reflected in increasing citations in the peer-reviewed literature.

The VAO is collaborating and cooperating with Observatories, missions, and new projects to integrate VO libraries into their processing environments to simplify and accelerate the dissemination of new data products. The VAO advises the community on the application of VO standards and libraries and on the use new technologies in developing applications. The VAO is supporting peer-reviewed journals in the development of the "interactive journal paper of the future," in which self-documenting data sets reported by the authors are curated and made accessible to readers through simple links. The VAO also shares its expertise with data archiving practitioners in other fields and is a resource in advising them on developing powerful tools and services that respond to their requirements.

With the VAO, the international astronomical community is at the forefront of multi-disciplinary activities in data and literature access and interoperability, such as those advocated by the U.S. Interagency Working Group on Digital Data in "*Harnessing the Power of Digital Data for Science and Society*," presented to the U.S. National Science and Technology Council (January 2009).[1] The challenges of managing a distributed organization, as described, for example, in "*Beyond Being There: A Blueprint for Advancing the Design, Development, and Evaluation of Virtual Organizations*"[2] and "*Cyberenvironment Project Management: Lessons Learned,*"[3] were taken into account in creating the VAO team and management structure.

## 2. ORGANIZATION

Direct management oversight of the VAO is provided by the Board of Directors of a limited liability company, Virtual Astronomical Observatory, LLC, which was established for the sole purpose of managing and operating the VAO. This company was established by the Associated Universities, Inc. (AUI) and the Associated Universities for Research in Astronomy, Inc. (AURA).

AUI, responsible for operating the National Radio Astronomy Observatory (NRAO) and for the North American interests in the Atacama Large Millimeter Array (ALMA), and AURA, responsible for operating the National Optical Astronomy Observatory (NOAO), the U.S. interest in the Gemini Observatory, the National Solar Observatory (NSO), and the Space Telescope Science Institute (STScI), together support the research interests of many U.S. astronomers and each has over 50 years of experience in managing national astronomical facilities for the community on behalf of NSF and/or NASA. Most of the institutions in our project team have ties to AURA or AUI. AUI and AURA vest oversight responsibility for the VAO in its own Board. Governance matters of the VAO, LLC are documented at http://www.usvao.org/governance/. Figure 1 shows the governance of the VAO.

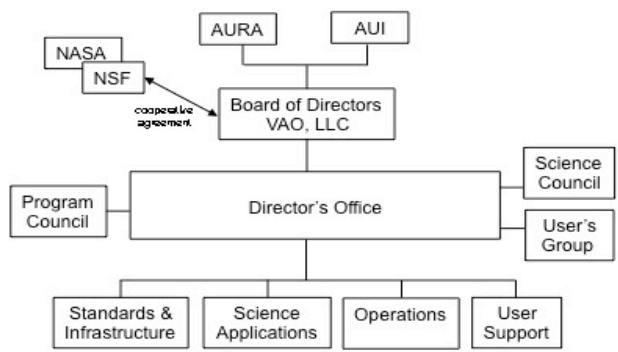

Figure 1. Organization of the U.S. Virtual Astronomical Observatory.

Specific details of the organization are provided in more detail below. In general, however, the VAO operates as a distributed organization and makes extensive use of modern communication methods to coordinate its activities on daily, weekly, monthly, and yearly time scales. Further, it involves the larger community in a variety of methods, including involvement in specific projects, providing advice on the implementation of VO standards and tools, and receiving advice on future scientific directions.

## 2.1 Board of Directors

The VAO Board of Directors has seven members. AUI and AURA each appoint three members, including the presidents of AUI (Dr. Ethan Schreier) and AURA (Dr. William Smith) who serve *ex officio*. Prof. Jay Gallagher (University of Wisconsin) has been appointed as the first chair; subsequent chairs will be selected by the other six board members.

### VAO, LLC Board of Directors

| | |
|---|---|
| Dr. Jay Gallagher (Chair) | University of Wisconsin |
| Dr. Bruce Margon | University of California, Santa Cruz |
| Dr. Catherine Pilachowski | Indiana University |
| Dr. Roscoe Giles | Boston University |
| Dr. Scott Kirkpatrick | Hebrew University, Jerusalem |
| Dr. Ethan Schreier (*ex officio*) | Associated Universities, Inc. (AUI) |
| Dr. William Smith (*ex officio*) | Associated Universities for Research in Astronomy, Inc. (AURA) |

Board members serve three-year terms, and the terms of the initial members are staggered so the board does not turn over completely at one time. The board's primary responsibility is to assure the overall scientific merit of the facility, and, through the review and approval of an annual Project Execution Plan prepared by the director, assure that resources are being used effectively and efficiently.

The board appoints a full-time director of the facility. The director is responsible for producing the Project Execution Plan for board approval, for the appointment of all senior personnel, and for all matters related to the conduct of the facility. The director is the primary, though not exclusive, contact with international VO development and support projects.

## 2.2 Senior Management and Staff

The VAO senior management team comprises the director's office staff (director, program manager, project scientist, technology advisor, business manager) and leads of the technical work areas. The senior management team meets by weekly telecon. The status of all work areas is reviewed regularly, with weekly status highlights and monthly reviews focusing on progress against the annual Project Execution Plan and monitoring of work against organizational deliverables. The program manager and work area leads support the director in preparing quarterly and annual reports. These reports cover all VAO-related activities, both NSF- and NASA-funded, and are intended to be responsive to both agencies. Reports are submitted through NSF's FastLane system.

Business administration services and general administrative support are provided through a subaward to AUI. There is a dedicated business manager who is responsible for the placement, monitoring, and reporting for all subawards and all financial dealings with NSF. The VAO, LLC, holds the funds that support the governance and management of the program (board, science council, and program council meetings, professional training and outreach events, etc.).

The staff in each work area also meet regularly by telecons convened by the work area leads. Discussions and decisions are documented. All work areas maintain work plans and schedules that are visible throughout the facility, and long-term tasks are broken into smaller, traceable components. Software products having distributed components or that are required to interoperate with other applications or components have well-defined interfaces. Work groups and project teams (which may cross work area boundaries) communicate by email and telecon discussions as needed, and conduct face-to-face meetings and work sessions.

## 2.3 Program Council

The VAO engages its collaborating organizations at two levels. The program council (PC) is composed of those senior managers at each core organization who have management and financial responsibility for the VAO work being done at that organization. The members of the PC work with the program manager and director to define work packages and allocate budgets for the annual Project Execution Plan. If the leaders of the various technical work areas are not satisfied with the quality or quantity of work being done by an organization, they come to the program manager, who then works with the PC to resolve the problem. Some members of the PC may also be themselves directly responsible for a VAO work area, but since most work areas involve several organizations, the PC member must be concerned with all tasks being done by his or her organization.

The organizations currently participating in the VAO are

- Associated Universities, Inc. (business management),
- Center for Advanced Computational Research, California Institute of Technology,
- High Energy Archive Science and Research Center, NASA Goddard Space Flight Center,
- Infrared Processing and Analysis Center, California Institute of Technology,
- Johns Hopkins University,
- National Center for Supercomputing Applications, University of Illinois,
- National Optical Astronomy Observatory,
- National Radio Astronomy Observatory,
- Smithsonian Astrophysical Observatory,
- Space Telescope Science Institute, and
- Jet Propulsion Laboratory, California Institute of Technology.

### 2.4 Science Council

The science council (SC) provides guidance to the director and the management team on matters of scientific priority. The SC is advisory to the director, who determines to what extent the SC's inputs can be addressed given available resources.

The SC meets at least once annually and has regular teleconferences and email communications, providing both formal and informal input. The SC Chair provides input and feedback on the annual Project Execution Plan in consultation with the SC. The VAO senior management team provides the SC with metrics showing the use and impact of the VAO for astronomical research.

### 2.5 Users Committee

The Users Committee is drawn from the scientific community and is composed of users and potential users of VAO science tools. Members include both those in the scientific community at large as well as with projects that the VAO is advising or with which the VAO is collaborating in the deployment of VO services and libraries. The Committee reports to the director and provides advice about the scientific functionality of existing VAO tools as well as recommendations about potential enhancements to those tools that would enable greater scientific productivity.

### 2.6 VAO Oversight Group

The NSF Division of Astronomical Sciences (AST) established a VAO Oversight Group (VOG) to provide oversight of the VAO and act as a point-of-contact NSF and NASA (the Agencies) and the project director. The VOG provides guidance in defining and refining the structure of the Project Execution Plan and approves annual updates to the document.

### 2.7 Team Meetings and Communications

The VAO facility team has face-to-face meetings at least twice per year. The primary purpose of these meetings is to assure program-wide awareness of progress and problems and develop the appropriate strategies for dealing with any issues. These meetings are coordinated with meetings of the science council, program council, and board of directors in order to economize on time and travel.

Effective management of this distributed facility requires efficient and innovative methods of communication. We make extensive use of telecons and a program-wide wiki site. We utilize communications technologies such as blogs, instant messaging, Internet-based videoconferences, and a dedicated VAO chat room.

## 2.8 External Collaborations

The VAO collaborating organizations include the major data archive centers for NASA's space astronomy missions and the major federally funded (NSF) ground-based national observatories, as well as prominent research universities and a national supercomputer facility. The VAO also has institutional associates with which it has agreed to collaborate and to include in activities such as team meetings, requirements analysis, prototyping, and evaluation and testing. These include the National Astronomy and Ionosphere Center/Arecibo Observatory, the Gemini Observatory, the Panoramic Survey Telescope And Rapid Response System (Pan-STARRS), and the Large Synoptic Survey Telescope (LSST); other such collaborations are in the process of being developed. We share information and experiences with VO-like initiatives in related fields, such as the NASA Planetary Data System and the NASA Virtual Solar-Terrestrial Observatory. The National Center for Supercomputing Applications (NCSA) at the University of Illinois is a core collaborator in the VAO, but also an external partner in terms of access to large-scale computational and storage resources. Microsoft Research is an ongoing external collaborator, with VO data access protocols fully integrated into the WorldWide Telescope (WWT) application. Through the VAO core organizations we have links to new facilities, such as the Jansky Very Large Array (VLA) and Atacama Large Millimeter/submillimeter Array (ALMA), and the VAO establishes contacts with other major development programs to discuss how they can benefit from take-up of VO standards and libraries and how the community can benefit by their provision of VO-compatible data discovery and access protocols.

The VAO recently had a community-wide call for collaborative proposals. As a result of that call, it is also advising four community groups on the incorporation of VO standards in their work or in methods by which their data can be more widely accessed. The four groups have projects that involve access to data and multi-wavelength models related to infrared observations of the Magellanic Clouds under the auspices of the *Spitzer* Surveying the Agents of Galaxy Evolution (SAGE) and *HERschel* Inventory of The Agents of Galaxy Evolution (HERITAGE) programs (PI: M. Meixner), access to data related to the Evolutionary Map of the Universe (EMU) and Variable and Slow Transient (VAST) Survey Science Programs of the Australian Square Kilometre Array Pathfinder (PI: T. Murphy), development of new software tools for the NASA Extragalactic Database (NED, PI: J. Schombert), and involve access to the databases of the American Association of Variable Star Observers (AAVSO, PI: M. Templeton). Future calls for collaborative proposals are anticipated.

## 2.9 Business and Financial Management

The Virtual Astronomical Observatory, LLC (Limited Liability Company), is a not-for-profit 501(c)(3) organization under the U.S. Internal Revenue Code and was registered in the District of Columbia on March 20, 2008. AUI and AURA are equal partners in the VAO, LLC, and have filed performance guarantee letters with NSF that assure "(a) the full and prompt payment and performance of all obligations, accrued and executory, which Contractor presently or hereafter may have to the Government under the Contract, and (b) the full and prompt payment and performance by Contractor of all other obligations and liabilities of Contractor to the Government, fixed or contingent, due or to become due, direct or indirect, now existing or hereafter and howsoever arising or incurred under the Contract." The full text of these letters is on file with the NSF Program Executive.

AUI and AURA have a formal operating agreement, signed by the presidents of these organizations on March 20, 2008, that defines the management and governance structure of the VAO. The principal place of business of the VAO, LLC, is 1400 16th Street NW, Suite 730, Washington, DC 20036.

The financial and business operations of the VAO, LLC, are managed through a subaward to AUI, which employs a business manager for this purpose. Specifically, the business manager is responsible for:

- Monitoring revenues and expenses of the VAO, LLC, account and reviewing the monthly general ledger;
- Managing daily operations of the accounts payable function and monitoring accounts receivable;
- Assisting in development of budgets and forecasts;
- Providing financial reports and analyses;
- Assisting with annual financial statement audits and A-133 compliance audits and tax filings;
- Managing all financial aspects of the Cooperative Agreement and subaward administration;
- Ensuring that NSF terms and conditions for the Cooperative Agreement are followed; and

- Analyzing costs listed on subaward invoices and determining allow-ability based on the U.S. Office of Management and Budget (OMB) regulations and resolving invoice problems.

The VAO, LLC, maintains a Financial and Administrative Policy and Procedures manual that documents all business and administrative operations. These policies and procedures are designed to assure that the VAO, LLC, is operated in full compliance with all relevant federal and state laws and with OMB Circulars A-110 (Uniform Administration Requirements for Grants and Agreements), A-122 (Cost Principles for Non-Profit Organizations), and A-133 (Audit of State and Local Governments and Other Non-Profit Institutions).

The policies and procedures defined in the manual include cash management, cash tracking, state and federal tax reporting, preparation and support for external audit, monthly financial statements, accounts payable, accounts receivable, general ledger, review of allowable and unallowable expenses, annual budget preparation, draw-down process, sub-recipient monitoring, and internal controls.

As the VAO, LLC, is a partnership between AUI and AURA, the VAO, LLC, financial and administrative policies and procedures are based on similar policies from these organizations and represent practices long recognized by both NSF and NASA.

## 2.10 Reporting

Quarterly reports are delivered to the NSF via FastLane. The Fourth Quarterly Report is submitted through the FastLane Annual Progress Report module, and conforms to NSF requirements on annual reports. Reports are submitted 30 days following the end of the reporting period.

# 3. TECHNICAL WORK

VAO technical work is organized into four primary areas, each with a technical lead reporting to the director's office, as shown in Figure 1.

1. **Standards and Infrastructure.** This work area encompasses all activities related to the development of standards and protocols and includes software development work related to reference implementations, support tools and services, and other aspects of the VO infrastructure such as interfaces to distributed storage for data sharing ("VOSpace"), data preservation methodologies, and security. Key elements of this work are to assure scalability and practical access to large-scale databases and archives and to assure timely adoption of new technologies. The work will include active outreach to data-producing organizations and projects (e.g., collaborating with LSST). Standards and Infrastructure has interfaces with Science Applications for applications integration and Operations for releases of software and infrastructure services to the community.

2. **Science Applications.** This work area includes development of innovative, general-purpose science applications that demonstrate the utilization of VO standards and infrastructure and identify where further development is needed. It also includes a program of community support for application development, e.g., with provision of an applications development toolkit. The VAO also helps to productize and disseminate tools developed by the community. The Science Applications work area has the primary responsibility for software testing.

3. **Operations.** This work area covers monitoring of VO services worldwide for aliveness and compliance with IVOA standards. Such work is essential to meet the goal of the VAO being a trusted and reliable means for the research community to discover and access data. The Operations group also has responsibility for maintaining the VAO software repository.

4. **User Support.** This work area encompasses the web site, documentation, news, event coordination, help desk services, community outreach/engagement, a user forum, and issue tracking.

## 4. ANNUAL PROGRAM PLAN

The VAO develops an annual program plan, officially called the Project Execution Plan (PEP), which describes the implementation, deployment, operations, and support pans for the next year and sets the context for future activities. The intent of the PEP is to

- Establish a common understanding with the funding agencies—NSF and NASA—of how the VAO evolves and meets their expectations for an operational virtual observatory that provides significant improvements to the research capability of the astronomical research community; and

- Provide a benchmark plan that defines responsibilities and informs all team members of the VAO's goals and deliverables.

The PEP is updated annually to reflect the changing nature of astronomical research, the changing nature of the computational environment in which astronomers do their work, the advice of the VAO science council, and the experience gained from working with the user community. It is responsive to the expectations identified by the funding agencies. The PEP is reviewed and approved by the VAO science council, VAO board of directors, and the NSF/NASA VAO Oversight Group. The final approved PEP is a public document available on the VAO website.

## 5. OUTREACH TO THE RESEARCH COMMUNITY

In addition to the Users Committee and Collaboration proposals (§2), the VAO undertakes outreach activities to the larger community, both to ensure that the community is aware of the emerging tools and standards as well as to provide another avenue for feedback on future directions. Examples of these outreach activities include a presence at the major astronomical meetings, VO Community Days, and social media.

The VAO has exhibited its tools and services at major astronomical meetings, most notably the American Astronomical Society's annual (winter) meeting (Figure 2). Past practice has been to sponsor an exhibit booth, which is open and staffed during the entire meeting. Particularly for the American Astronomical Society (AAS) meetings, there is attendance from a significant fraction of the U.S. astronomical community, as well as from the international community, allowing ample opportunity for demonstrations and extended discussions. At the most recent AAS meeting, the VAO also sponsored a workshop, "Science Tools for Data-Intensive Astronomy," and a Special Session, "Cyber-Discovery and Science for the Decade."

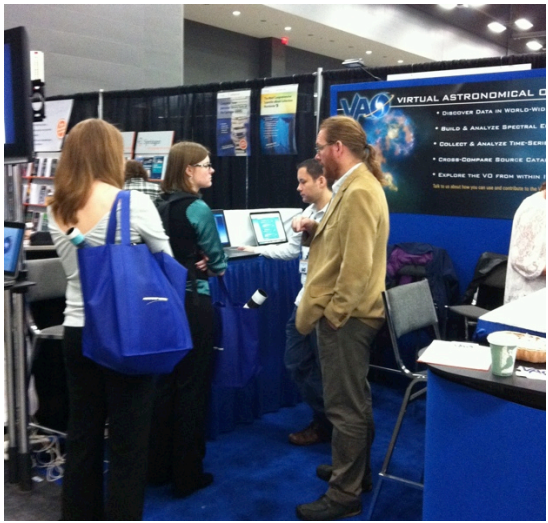 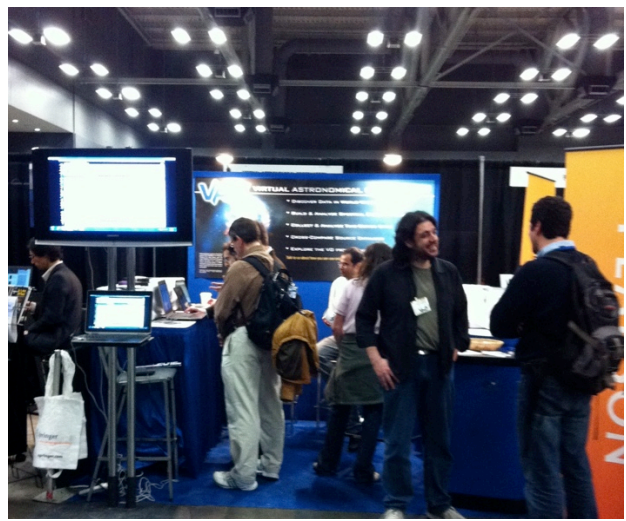

Figure 2. Virtual Astronomy Observatory exhibit at the American Astronomical Society 219th Meeting.

The VAO has begun to organize and sponsor VO Community Days. These are typically half-day sessions at an institution during which time VO tools and services are demonstrated; hands-on sessions in which participants use VAO tools to address specific science questions are being introduced as well. To date, VO Community Days have been held at Cambridge, MA (2011 November 30), Pasadena, CA (2011 December 6), and Tucson, AZ (2012 March 9); additional VO Community Days are being planned. The VAO is also beginning to make similar presentations and conducted demonstrations at community-organized summer schools, with the first example being at the Summer School in Statistics for Astronomers VIII held at the Pennsylvania State University.

Finally, recognizing that an increasing number of astronomers, particularly younger members of the community, are heavily involved in social media outlets, the VAO has a presence on many of the standard social media outlets such as Facebook (Figure 3) and Twitter. Through updates and posts, the VAO provides updates on software updates, papers making use of VO tools and services, upcoming events, and other topics of potential interest.

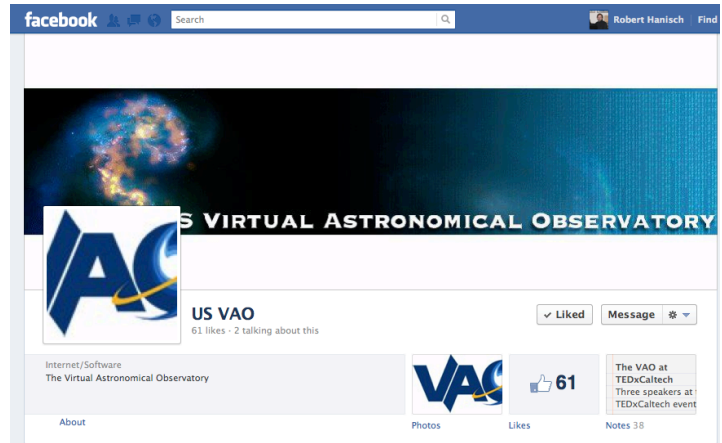

Figure 3. Facebook page for the Virtual Astronomical Observatory.

## 6. CONCLUSIONS

The management of a distributed organization such as the VAO, which is a combination of development and operations activities and which is supported by two independent agencies (with different modes of funding), presents some special challenges. Moreover, the VAO operates as part of a larger virtual organization, the International Virtual Observatory Alliance, which has no central funding at all but rather relies on the mutual cooperation of its members.

In order that the VAO would be seen as an entity that is of and for the research community, a dedicated not-for-profit company was established to manage the governance and business functions. Thus, no one organization "owns" or has special influence over the VAO, and its board of directors represents the community's interests, nor is the VAO "lost" among the large (in size and budget) physical observatories managed by organizations such as AUI and AURA. The VAO has benefitted strongly from having a board that is dedicated to its oversight.

Each of the organizations participating in the VAO collaboration brings unique skills to the effort. In a few cases responsibility for certain tasks or development activities lies fully within one organization, but more often the expertise from two or more groups is required to develop or support some capability. Because the work is so strongly collaborative we rely heavily on electronic communications, from e-mail and telecons to wiki sites, shared software repositories, and Internet-based e-meetings. In our experience, however, nothing replaces the need for face-to-face discussions, particularly for technical work sessions (splinter groups, tiger teams) and project-wide team meetings.

There are costs associated with managing a distributed project that do not pertain to a group working all within the same organization and under a single management structure. One has to make a concerted effort on communication, as casual interactions (having a chat with a colleague across the hall) are more difficult. Modern communication technologies such as wikis and chat rooms and e-meetings help, and will perhaps become even more widely used as more organizations adopt a flexible workplace.

## ACKNOWLEDGMENTS

This paper describes work done with the support of the US Virtual Astronomical Observatory. The VAO is jointly funded by the National Science Foundation (under Cooperative Agreement AST-0834235) and by the National Aeronautics and Space Administration. The VAO is managed by the VAO, LLC, a non-profit 501(c)(3) organization registered in the District of Columbia and a collaborative effort of the Association of Universities for Research in Astronomy (AURA) and the Associated Universities, Inc. (AUI). Part of this research was carried out at the Jet Propulsion Laboratory, California Institute of Technology, under a contract with the National Aeronautics and Space Administration.